\documentclass[conference]{IEEEtran}

\usepackage[utf8]{inputenc}
\usepackage[T1]{fontenc}
\usepackage{amsmath}
\usepackage{amssymb}
\usepackage{graphicx}
\graphicspath{{figures/}}
\usepackage{subcaption}
\usepackage{threeparttable}
\usepackage{comment}
\usepackage{cite}
\usepackage{url}
\usepackage{xcolor}
\usepackage{hyperref}

\definecolor{mayapink}{HTML}{FF1493}

\renewcommand{\phi}{\varphi} 

\title{Power Scaling Across Link Utilization with GPAM-4 Signaling and Event-Driven Reception}

\author{
\IEEEauthorblockN{Maya~Mekel, Jeries Mattar, and Nicolás Wainstein}
\IEEEauthorblockA{Andrew and Erna Viterbi Faculty of Electrical and Computer Engineering,
Technion – Israel Institute of Technology, Israel} 
Corresponding authors: maya.mekel@campus.technion.ac.il, nico@technion.ac.il
}

\begin{document}

\maketitle


\begin{abstract}
This work highlights the potential of combining event-based signaling with event-driven reception to achieve receiver power scaling across link utilization, without requiring wake-up, locking, or data-rate reconfiguration. The approach is demonstrated using the proposed four-level Gaussian pulse-amplitude modulation (GPAM-4) scheme and a corresponding event-driven receiver. By employing a resettable ring voltage-controlled oscillator (RRVCO) as its front end, the receiver achieves inherently event-driven symbol measurement, initiated directly by each received pulse. Symbols are decoded directly from the final RRVCO phase state, eliminating a counter-based back end. Simulated in 28-nm CMOS, the receiver power is evaluated across a \(100\times\) range of effective data rates---from 0.15~mW at the peak rate of 8~Gb/s, down to 2.5~\(\mu\)W under active/idle operation and 1.6~\(\mu\)W under data rate change at 80~Mb/s. The receiver thus achieves power reductions of \(59.5\times\) and \(90.0\times\), while its energy per bit increases by only \(1.68\times\) and \(1.11\times\), respectively.
\end{abstract}



\section{Introduction}
\label{sec:introduction}

Link utilization in high-speed wireline SerDes is determined by both the configured data rate and the fraction of time the link is actively transmitting, with periods of no data transmission referred to as idle periods. However, receiver power often scales weakly with link utilization due to continuously operating clocking, equalization, and front-end circuits, creating an energy-proportionality gap. This motivates receivers whose power consumption scales with link utilization \cite{Kim2025ROOCDR, Anand2015EmbeddedClock, Kim2020ROODFE, Shekhar2019Bidirectional, Choi2015BurstMode, Smith2021DEMS, Smith2023DEMS}.

Prior work has addressed this by adapting conventional clocked receivers for burst-mode operation and data-rate reconfiguration, requiring additional control mechanisms that introduce wake-up, locking, or reconfiguration overhead \cite{Kim2025ROOCDR, Anand2015EmbeddedClock, Kim2020ROODFE, Shekhar2019Bidirectional, Choi2015BurstMode}.
An alternative architectural direction is to exploit event-driven reception, in which incoming signal events initiate receiver operation rather than a free-running clock, potentially enabling power scaling across both dimensions of link utilization without explicit mode-transition control or data-rate reconfiguration.
Such reception can be enabled by co-designing the receiver with a clock-encoded modulation scheme, in which timing information is carried by the signal itself.

Event-driven operation has been demonstrated in analog-to-digital converters \cite{Naaman2024IFTEMHardware,Teixeira2023IntegrateFire}, event-driven sensing \cite{Llopart2022Timepix4,Maj2026ThreeFI}, neuromorphic circuits \cite{Granizo2023LIF}, and communication links across physical media \cite{Crepaldi2014PulsedReceiver,Li2025OpticalReceiver,Qiao2019AERLink,Smith2023DEMS}.
Specifically, event-driven reception has been demonstrated in wireless and optical links using pulse-based signaling \cite{Crepaldi2014PulsedReceiver,Li2025OpticalReceiver}, and in wireline links using address-event signaling for neuromorphic communication \cite{ZamarrenoRamos2011AER,Qiao2019AERLink} and differential edge-modulated signaling (DEMS) for high-speed digital communication \cite{Smith2021DEMS,Smith2023DEMS}.
Among these works, DEMS demonstrates event-driven reception for high-speed digital wireline communication, supporting data-rate changes without receiver reconfiguration. However, it does not demonstrate power scaling with active/idle utilization.
Data-driven synchronization \cite{Ha2013GRO} differs from event-driven reception: the former relies on established synchronization, whereas in the latter each symbol initiates decoding independently.

This work proposes explicitly exploiting the connection between event-driven reception and receiver power scaling across both dimensions of link utilization. The approach is demonstrated through 28-nm CMOS simulations of a proposed event-driven receiver based on a resettable ring voltage-controlled oscillator (RRVCO), co-designed with four-level Gaussian pulse-amplitude modulation (GPAM-4) and receiver-aware symbol amplitude selection.
Extending beyond the demonstrated capabilities of the prior DEMS receiver, this work demonstrates receiver power scaling with active/idle utilization in addition to data-rate changes.
Unlike VCO-based ADCs, the RRVCO decodes each symbol directly from its final state, eliminating the counter and digital decoding back-end.


\section{Proposed Signaling Modulation Scheme}
\label{sec:modulation_scheme}

\begin{figure}[t]
    \centering
    \subfloat[]{\includegraphics[trim=0 3.65pt 0 0, clip]{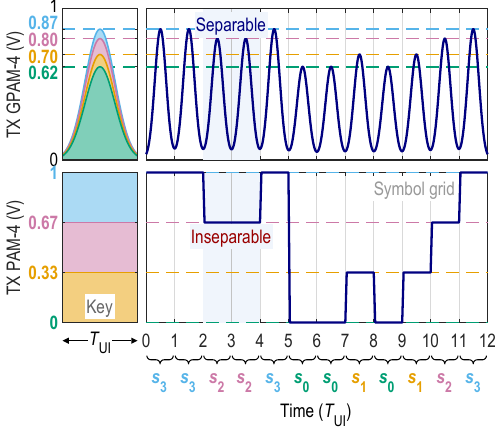}\label{fig:pam4_gpam4_waveforms}}\\[2pt]
    \subfloat[]{\includegraphics[trim=0 3pt 0 0, clip]{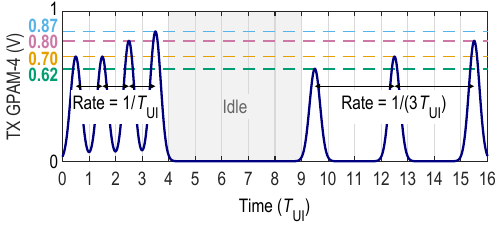}\label{fig:gpam4_nonuniform}}
\caption{(a) PAM-4 vs. GPAM-4 for the same symbol sequence. (b) GPAM-4 bursts at different rates separated by an idle period.}
    \label{fig:waveforms}
    \vspace*{-1em}
\end{figure}

At the transmitter, symbols are formed within intervals of duration $T_{\mathrm{UI}}$. Event-driven reception requires each symbol to produce a detectable event, for which a pulse can serve. The proposed GPAM-4 modulation adopts Gaussian pulses $p(t;\sigma)=\exp(-t^2/2\sigma^2)$ for their fast spectral decay, reducing high-frequency content and potentially improving crosstalk robustness. Each pulse is centered within its interval and scaled by one of four amplitude levels $a_0,~a_1,~a_2,~a_3$ (see Section~\ref{sec:amp_selection}).
Since Gaussian pulses have infinite temporal support, $\sigma$ is chosen as a fraction of $T_{\mathrm{UI}}$ to keep each pulse effectively localized within its symbol interval.

Fig.~\ref{fig:pam4_gpam4_waveforms} compares GPAM-4 and PAM-4 for the same symbol sequence and unit interval ($T_{\mathrm{UI}}$).
Unlike PAM-4, whose symbol boundaries follow a fixed time grid, GPAM-4 pulses define individually observable symbol events.
As illustrated in Fig.~\ref{fig:gpam4_nonuniform}, varying pulse spacing accommodates changes in data rate, while the absence of pulses represents idle periods. Exploiting this event-based signaling property through event-driven reception is key to enabling receiver power scaling across both dimensions of link utilization.


\section{Receiver Architecture}
\label{sec:receiver_architecture}


\begin{figure}[t]
    \centering
    \subfloat[]{\includegraphics{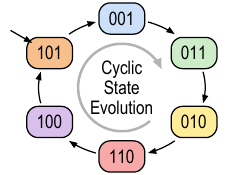}%
        \label{fig:state_evolution}}%
    \hfill
    \subfloat[]{\includegraphics{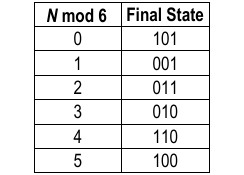}%
        \label{fig:toggle_mapping}}
    \caption{Three-stage ring VCO states.
(a)~Cyclic state evolution.
(b)~Final state versus toggle count for an RRVCO initialized to 101.}
    \label{fig:rrvco_states}
    \vspace{-1em}
\end{figure}

In this work, a ring-VCO front end is employed, allowing each pulse to directly initiate oscillation and measurement without prior detection, unlike event-driven receivers requiring explicit pulse detection in advance \cite{Maj2026ThreeFI, Jeon2020PeakTriggered, DiFrancesco2016TOFPET2, Achtenberg2020PeakReview}. Compared with comparator-bank receivers \cite{WeltinWu2013LevelCrossing}, which may incur significant static power, its predominantly dynamic power consumption may offer an advantage during idle periods.

Consider an $n$-stage ring VCO with $n$ phases, whose instantaneous logic levels collectively define its state. The VCO integrates the input voltage into accumulated phase, with each fixed phase increment triggering the next sequential phase toggle, advancing the VCO through a finite cyclic sequence of states. Higher input voltages increase the phase accumulation rate and thus the toggling frequency.

The symbol-decoding principle of this work is inspired by VCO-based ADCs, which recover the input amplitude by counting phase toggles over a predefined window, typically requiring counter- or phase-difference-based back-end processing \cite{ParkThesis2009}. For digital communication, however, only a few symbols must be distinguished: owing to the correspondence between phase toggling and state progression, an $n$-stage ring VCO with a known initial state distinguishes $2n$ toggle counts directly from its state, without a counter. The proposed receiver establishes this known initial state for each symbol using a RRVCO, which can be reset to a predefined state through control signals. Previously demonstrated in VCO-based ADCs \cite{Lu2026RVCO}, the RRVCO is here employed for symbol decoding.
Understanding the operation of the receiver requires establishing two aspects: the response each scaled pulse induces in the RRVCO, and how consecutive pulses are handled.

\subsection{Receiver Response to Scaled Pulses}
\label{subsec:single_pulse_response}

The receiver is built around a three-stage RRVCO with $101$ as its reset state.
Under the effective pulse localization established in Section~\ref{sec:modulation_scheme}, a Gaussian pulse of amplitude $a$ is approximated as a return-to-zero (RZ) pulse.
Such a pulse limits RRVCO phase accumulation to a finite interval, inducing a finite number of toggles, denoted by $N_a$. Starting from the reset state, the RRVCO advances through the cyclic state sequence illustrated in Fig.~\ref{fig:state_evolution}, with its final state determined by $N_a$ through the mapping in Fig.~\ref{fig:toggle_mapping}.
For GPAM-4 pulses, the relationship between $a$ and $N_a$ is characterized in Section~\ref{sec:amp_selection}, where it guides the selection of the four amplitudes.

\subsection{Receiver Operation Overview}
\label{subsec:receiver_operation_overview}

\begin{figure}[t]
    \centering
    \subfloat[]{\includegraphics[trim={0.1cm 0.1cm 0.1cm 0.1cm},clip]{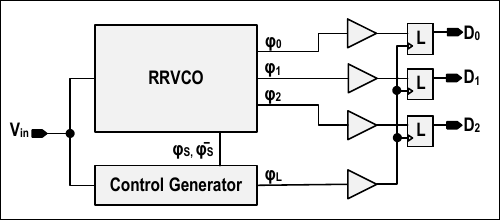}%
\label{fig:receiver_architecture_block}}\\[0pt]
    \subfloat[]{\includegraphics[trim={0.1cm 5pt 0.1cm 0.1cm},clip]{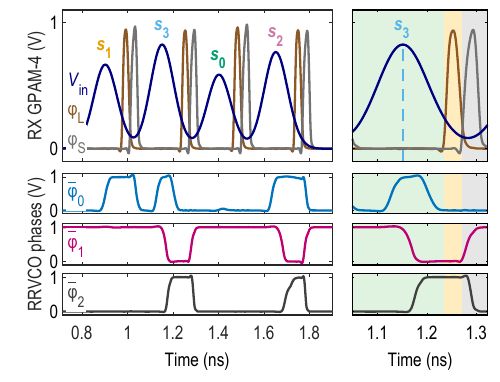}%
        \label{fig:waveform_decodings}}
\caption{(a) Receiver architecture (keepers omitted for clarity). (b) Receiver waveforms for consecutive GPAM-4 symbols. Symbol amplitudes and expected toggle counts follow the selection in Fig.~\ref{fig:map_selected}. Complemented RRVCO phases $\bar{\phi}_0$--$\bar{\phi}_2$ are advanced by 16~ps to compensate for inverter delay. Zoom: green shading highlights pulse-induced phase toggling, yellow shading highlights stable phases during latching, and gray shading indicates reset.
}
    \vspace*{-1em}
\end{figure}

Building on the RRVCO response to an individual scaled pulse, the remaining receiver circuitry (Fig.~\ref{fig:receiver_architecture_block}) captures the final state induced by each pulse and then resets the RRVCO for the next one.
Each symbol $s_i$ arrives as a pulse in $V_{\mathrm{in}}$ scaled by $a_i$, driving both the RRVCO and a self-timed control path. The pulse induces $N_{a_i}$ toggles, after which the RRVCO settles in its final state. 
The control path detects when the falling edge of the pulse crosses the threshold $V_{\mathrm{th},L}$ and asserts $\phi_L$ after a delay $T_{d,L}$.
The RRVCO phases $\phi_0$, $\phi_1$, and $\phi_2$, together with $\phi_L$, are buffered and fed to true single-phase clock latches \cite{Razavi2016TSPC}, which capture the final phase state as $[D_0,D_1,D_2]$ on the falling edge of the buffered $\phi_L$. Correct capture requires the phases to remain unchanged while the latches are transparent for a minimum duration $T_{\mathrm{latch}}$. Symbols with $N_{a_i}=0$ are also decoded: as long as the pulse activates the control path, the reset state $101$ is captured as a valid symbol. 

To ensure capture completes before reset, the control path asserts the complementary reset signals $\phi_S$ and $\overline{\phi_S}$ after the falling edge of $\phi_L$ crosses $V_{\mathrm{th},S}$, with a detection delay $T_{d,S}$. The combined assertion interval of $\phi_S$ and $\overline{\phi_S}$ spans at least $T_{\mathrm{reset}}$, the minimum duration required to fully reset the RRVCO. Keepers at the buffered phase nodes prevent state drift during idle periods.
Fig.~\ref{fig:waveform_decodings} illustrates the receiver operation. The phase waveforms show the toggles induced by each pulse, the stable phase states during capture, and their restoration to $101$ during reset.

Since each received pulse triggers its own measurement, capture, and reset, the receiver is event-driven: it requires no continuously running clock, transitions from idle to active without wake-up or locking, and accommodates changes in symbol arrival rate without receiver reconfiguration.

\subsection{Resettable Ring VCO Implementation}
\label{subsec:rrvco}

\begin{figure}[t]
    \centering   \includegraphics[trim={0.1cm 0.05cm 0.1cm 0.025cm},clip]{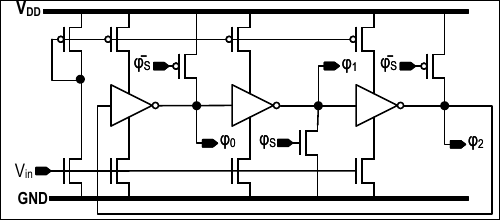}
    \caption{Implementation of the proposed three-stage
    resettable ring VCO.}
    \label{fig:rrvco}
    \vspace*{-1em}
\end{figure}

The RRVCO comprises a three-stage current-starved ring oscillator with phases $\phi_0$, $\phi_1$, and $\phi_2$, with additional reset transistors driven by $\phi_S$ and $\overline{\phi_S}$, which force the initial state $[\phi_0,\phi_1,\phi_2]=[1,0,1]$ (Fig.~\ref{fig:rrvco}). NMOS current-starving devices are driven by $V_{\mathrm{in}}$, while a current-mirror bias sets the PMOS devices to track the input-controlled pull-down current.


\section{GPAM-4 Amplitude Selection Methodology}
\label{sec:amp_selection}

\begin{figure}[t]
    \centering
    \subfloat[]{\includegraphics[width=0.9\columnwidth, trim=0 8pt 0 0, clip]{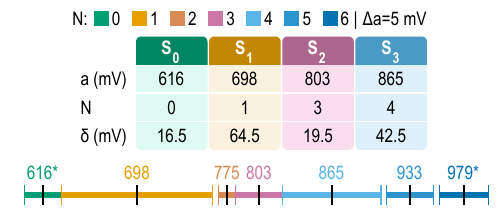}%
        \label{fig:map_selected}}\\[-1pt]
    \subfloat[]{\includegraphics[width=0.9\columnwidth, trim=0 8pt 0 3pt, clip]{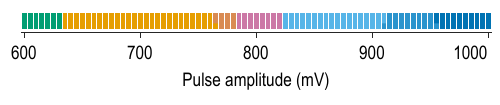}%
        \label{fig:map_bare}}
\caption{Amplitude selection and amplitude-to-$N$ map under nominal conditions (Section~\ref{sec:performance}). (a)~Ticks mark the centers of stable regions (bars); labels identify the four selected symbols; * marks regions bounded by the sweep limit. (b)~Map; cells split by the proportions of pulses yielding different $N$.}
    \label{fig:GPAM4_map}
    \vspace*{-1em}
\end{figure}

Symbol amplitudes are selected according to the receiver response to scaled pulses (Section~\ref{subsec:single_pulse_response}). They must induce distinct final states, and should preferably also maximize margin against variations in the receiver response while minimizing RRVCO toggling, and hence dynamic power. Since the latter depends on $N_a$, the selection is based on a pulse-amplitude-to-$N$ map, constructed by sweeping $a$ over the available range with a finite resolution $\Delta a$, with the channel included, and recording the final state captured for each pulse. 

Since distinct values of $N_a$ may share a final state (Fig.~\ref{fig:toggle_mapping}), absolute $N_a$ is recovered from its monotonic increase with $a$. 
Fig.~\ref{fig:map_bare} shows a map derived under nominal conditions (Section~\ref{sec:performance}). 
Amplitudes at which all pulses induce the same $N_a$ form stable regions, while amplitudes at which pulses induce different counts mark transitions between them. Stable regions may have unequal widths, as illustrated by the narrow $N=2$ region. The supported modulation order is limited by the number of distinct final states reachable over the available amplitude range. When stable regions outnumber the required levels, the preferences can be applied. Fig.~\ref{fig:map_selected} labels the selected amplitudes and the distance $\delta_a$ of each from its nearest transition. Owing to the finite sweep resolution, region boundaries, and hence $\delta_a$, are known to within $\Delta a$.

In conventional PAM-4, the observable is the signal amplitude itself, so uniform amplitude spacing maximizes separation, and eye-diagram height indicates margin. Here, the observable is the final state, so the spacing is dictated by the map, and $\delta_a$ serves as the counterpart of eye height.


\section{Performance}
\label{sec:performance}

The receiver is designed and simulated in Cadence Virtuoso using a 28-nm CMOS process, with an ideal series-terminated transmitter driving an electromagnetic model of an on-die 2.5-mm interposer routing with width $2\,\mu\mathrm{m}$, terminated by a $30\,\mathrm{fF}$ capacitor that models the bump and ESD capacitance. 
Unless otherwise noted, simulations use nominal conditions (typical corner, $V_{\mathrm{DD}}=1$ V, $T=70\,^{\circ}$C) and a GPAM-4-modulated PRBS7Q sequence (a 127-symbol Gray-coded PAM-4 pattern derived from PRBS7) with $\sigma = T_{\mathrm{UI}}/5$, $T_{\mathrm{UI}}=250$ ps (setting the peak data rate), and amplitudes selected in Fig.~\ref{fig:map_selected}.

\subsection{Power Scaling with Both Dimensions of Link Utilization}
\label{subsec:power_scaling}

\begin{figure}[t]
    \centering
    \includegraphics[width=\columnwidth]{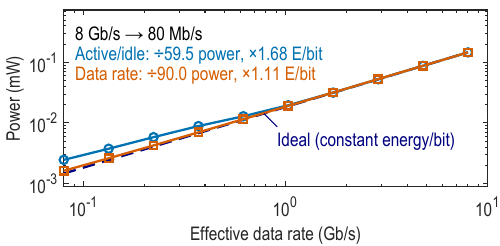}
\caption{Simulated receiver power vs.\ effective data rate for active/idle and data-rate sweeps; markers show simulated points, interpolated by the curves.}
\label{fig:power_scaling}
    \vspace*{-1em}
\end{figure}

Receiver power is calculated from simulated supply current, and energy-efficiency degradation is measured relative to the energy per bit at peak data rate, the ideal-proportionality reference. The effective data rate is transmitted bits per total time. For active/idle operation, the receiver is driven by burst--idle cycles, each a peak-rate GPAM-4 burst followed by an idle interval of swept duration. For changing-data-rate operation, the same sequence is applied with swept uniform spacing, so that the effective data rate equals the data rate.

As the effective data rate drops $100\times$, from 8~Gb/s to 80~Mb/s, receiver power scales with both dimensions (Fig.~\ref{fig:power_scaling}): it decreases by $59.5\times$ and $90.0\times$ for the active/idle and data-rate sweeps, while energy per bit increases by $1.68\times$ and $1.11\times$, respectively. The difference is attributed to static current during long idle intervals.

\subsection{GPAM-4 Signal Integrity and Pulse-Width Dependence}
\label{subsec:PVT_section}

\begin{figure}[t]
    \centering
    \subfloat[]{\includegraphics[width=0.9\columnwidth]{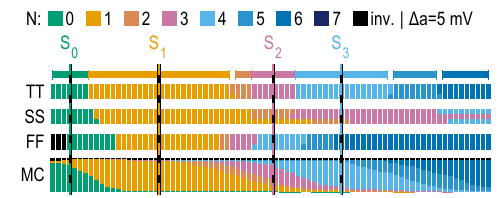}%
        \label{fig:corner_Nmap}}\\[-1pt]
       \subfloat[]{\includegraphics[width=0.9\columnwidth, trim=0 6pt 0 0, clip]{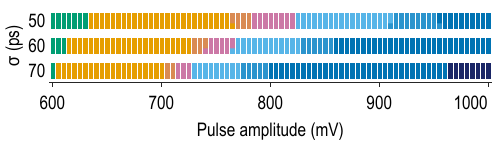}%
        \label{fig:sigma_Nmap}}
        \vspace{-0.4em}
\caption{Amplitude-to-$N$ maps; inv.\ marks undetected pulses. (a)~TT (1~V, 70~$^\circ$C) with the nominal stable regions (bars) and selected amplitudes $s_0$--$s_3$ (dashed), SS (0.9~V, 100~$^\circ$C), and FF (1.1~V, 0~$^\circ$C) corners, and local mismatch at TT (100 Monte Carlo runs). (b)~TT for different $\sigma = T_{\mathrm{UI}}/5$.}
    \label{fig:map_sensitivity}
\end{figure}



To assess signal integrity, the pulse-amplitude-to-$N$ map is constructed at TT, SS, FF, and under local mismatch at TT (Fig.~\ref{fig:corner_Nmap}). Stable-region boundaries shift nonuniformly, requiring level-wise rather than global-offset amplitude calibration. The PVT analysis reveals that four-symbol encoding is infeasible at SS under the considered pulse parameters. 
Under local mismatch, the yield for four symbols with an amplitude margin of at least 10 mV is 85/100. This motivates per-die calibration, left for future work, in which a $\Delta a$-step amplitude staircase is transmitted within a PRBS or a representative transaction, and receiver feedback selects the four amplitudes.
Recalibrating the TT amplitudes to the per-die selections (largest margin first, then lowest $N$) requires a trim range of 217.5 mV, dominated by remapping, since only 8 dies retain the TT set $\{0,1,3,4\}$. The required resolution of 1.25 mV is bounded by the $\Delta a = 5$ mV sweep grid.

Fig.~\ref{fig:sigma_Nmap} shows the TT map for different pulse widths. A larger $\sigma$ induces more toggles at a given amplitude, shifting the stable regions toward lower amplitudes.

\subsection{Comparison with Prior Work}
\label{subsec:comparison}
Table~\ref{tab:comparison} compares the proposed receiver with prior power-scaling receivers, with metrics extracted for the receiver only. Unlike the others, it demonstrates active/idle operation without wake-up, locking, or rate reconfiguration.

\begin{table}[t]
\centering
\caption{Comparison with Prior Work}
\label{tab:comparison}
\begin{threeparttable}
\setlength{\tabcolsep}{2pt}
\begin{tabular*}{\columnwidth}{@{\extracolsep{\fill}}lccccc@{}}
\hline
\noalign{\vskip 2pt}
& \textbf{This work}\tnote{$\dagger$}
& \shortstack{SSCL\\ `21\cite{Smith2021DEMS}}
& \shortstack{TCAS-II\\ `25\cite{Kim2025ROOCDR}}
& \shortstack{JSSC\\ `20\cite{Kim2020ROODFE}}
& \shortstack{CICC\\ `25\cite{Choi2025DieToDie}} \\
\hline
\noalign{\vskip 1pt}
Technology (nm)
& 28 & 65 & 28 & 65 & 28 \\

Supply (V)
& 1 & n.r. & n.r. & 1.1 & n.r. \\

Data Rate / Range (Gb/s)
& Up to 8 & 3--10 & 2.5--48 & 12 & 28 \\

Active/Idle Demonstrated
& Yes & n.r. & Yes & Yes & Yes \\

Rate Reconfig.~Required
& No & No & Yes & n.a.\tnote{b} & n.a.\tnote{b} \\

Wake-Up / Lock Time (ns)
& None\tnote{a} & None\tnote{a} & 16.6 & 10 & 3.11\tnote{c} \\

Channel Loss (dB)
& 1.3\tnote{d} & 10 & 7.1 & 20 & 6.06 \\

Peak-Rate Energy (fJ/bit)
& 18 & 810 & 1400 & 3800 & 140 \\
\hline
\end{tabular*}
\begin{tablenotes}[flushleft]
\scriptsize
\item[] n.r.: not reported; n.a.: not applicable. $^\dagger$ Simulated; others measured.
\item[a] Self-synchronous; no locking or timing reconfiguration.
\item[b] Fixed-rate implementation. \textsuperscript{c} Not distinguished from pipeline latency.
\item[d] GPAM-4 channel loss evaluated at $1/T_{\mathrm{UI}}=4$ GHz.
\end{tablenotes}
\end{threeparttable}
\vspace*{-1.5em}
\end{table}


\section{Conclusion}
This work highlights the potential of combining event-based signaling with event-driven reception to achieve receiver power scaling across link utilization without requiring wake-up, locking, or data-rate reconfiguration. The approach is demonstrated using GPAM-4 signaling and an RRVCO-based receiver, in which each received pulse carries a symbol and directly initiates its measurement, capture, and reset. Future work will focus on improving the robustness of the proposed implementation across PVT through calibration and extending the framework to other return-to-zero (RZ) pulse shapes.


\section*{Acknowledgments}
This work was funded by the Israel Innovation Authority no. 91302. N. Wainstein is the Robert J. Shillman Fellow. 

\bibliographystyle{IEEEtran}
\bibliography{refs}

\end{document}